\documentclass{article}

\usepackage[usenames,dvipsnames,svgnames]{xcolor}

\usepackage{hyperref}
\hypersetup{
    colorlinks=true,
    citecolor=MidnightBlue,
    urlcolor=Bittersweet,
}

\usepackage[margin=1in]{geometry}

\usepackage{cite}
\usepackage{amsmath,amssymb,amsfonts}
\usepackage{algorithmic}
\usepackage{graphicx}
\usepackage{textcomp}
\usepackage{algorithm}
\usepackage{booktabs}
\usepackage{tikz}
\usepackage{bm}
\usepackage{subcaption}
\usetikzlibrary{spy}
\newcommand{\param}{\boldsymbol{\theta}}

\usepackage{xcolor}
\usepackage{cite}
\usepackage{amsmath,amssymb,amsfonts}
\usepackage{algorithmic}
\usepackage{graphicx}
\usepackage{textcomp}
\usepackage{algorithm}
\usepackage{booktabs}
\usepackage{tikz}
\newcommand{\BLUE}{\textcolor{black}}

\newcommand{\x}{\mathbf{x}}
\newcommand{\y}{\mathbf{y}}
\newcommand{\X}{\mathbf{X}}
\newcommand{\Z}{\mathbf{Z}}
\newcommand{\omg}{\mathbf{\Omega}}

\newcommand{\R}{\mathbb{R}}
\newcommand*\samethanks[1][\value{footnote}]{\footnotemark[#1]}

\begin{document}
\title{Enhancing Low-dose CT Image Reconstruction by Integrating Supervised and Unsupervised Learning}
\author{Ling Chen\thanks{L. Chen and Y. Long are with the University of Michigan Shanghai Jiao Tong University Joint Institute, Shanghai Jiao Tong University, Shanghai 200240, China (chen ling@sjtu.edu.cn; yong.long@sjtu.edu.cn). }\, , 
Zhishen Huang\thanks{Z. Huang is with Amazon Inc., Seattle, WA 98109, USA (zhishen.huang@colorado.edu). This work was performed when Z.Huang was with Michigan State University.}\, , 
Yong Long\samethanks[1]
\,and Saiprasad Ravishankar\thanks{S. Ravishankar is with the Department of Computational Mathematics, Science and Engineering and the Department of Biomedical Engineering, Michigan State University, East Lansing, MI 48824, USA (ravisha3@msu.edu).}
}

\maketitle

\begin{abstract}
Traditional model-based image reconstruction (MBIR) methods combine forward and noise models with simple object priors. Recent application of deep learning methods for image reconstruction provides a successful data-driven approach to addressing the challenges when reconstructing images with undersampled measurements or various types of noise. In this work, we propose a hybrid supervised-unsupervised learning framework for X-ray computed tomography (CT) image reconstruction. The proposed learning formulation leverages both sparsity or unsupervised learning-based priors and neural network reconstructors to simulate a fixed-point iteration process. Each proposed trained block consists of a deterministic MBIR solver and a neural network. The information flows in parallel through these two reconstructors and is then optimally combined. Multiple such blocks are cascaded to form a reconstruction pipeline. We demonstrate the efficacy of this learned hybrid model for low-dose CT image reconstruction with limited training data, where we use the NIH AAPM Mayo Clinic Low Dose CT Grand Challenge dataset for training and testing. In our experiments, we study combinations of supervised deep network reconstructors and MBIR solver with learned sparse representation-based priors or analytical priors. Our results demonstrate the promising performance of the proposed framework compared to recent low-dose CT reconstruction methods.
\end{abstract}

\section{Introduction}
\label{sec:introduction}

X-ray computed tomography (CT) is a non-invasive imaging technique widely used in modern clinical diagnosis 
for visualizing organs and tissues. Early commercialized CT methods used into early 2010s included algebraic reconstruction, and filtered back projection (FBP)~\cite{1984Practical}, where measurements of X-ray flux are collected from multiple views or angles as the \textit{sinogram}, and such projections are transformed back into readable images with inverse Radon transform with a high-pass filter. The FBP method is susceptible to noise 
and as a byproduct also suffers from low-contrast lesion details. The growth of computational power led to the advent of iterative reconstruction (IR) methods for CT, which leverage models of the imaging physics and noise statistics and
prior domain knowledge about the objects being scanned
and incorporate regularization terms in a penalized least-square formulation of the image reconstruction problem. 
IR methods reduce noise compared to FBP, and with the resulting enhanced image quality, a reduction in the X-ray dose applied on patients becomes feasible. Edge-preserving (EP) regularization~\cite{EP_97} assumes the image is approximately sparse in the gradient domain. Dictionary learning-based methods~\cite{xu:12:ldx} provide improved image reconstruction quality compared to non-adaptive model-based IR schemes, but incur high computational cost for sparse encoding. 
Penalized weighted least squares (PWLS) methods with regularizers involving learned sparsifying transforms (PWLS-ST~\cite{2017Low}) or a union of learned transforms (PWLS-ULTRA~\cite{zheng2018,MBIR_review_21}) leverage both computational efficiency (cheap sparse coding in transform domain) and the representation power of learned models (transforms). 
Efficient optimization methods \cite{2012A, 2013First, Donghwan2014Combining } have been studied to solve MBIR optimization problems with particular regularizers.


Low-dose CT (LDCT) aims to lessen patients' exposure to ionizing radiation, however, the performance of IR methods in the LDCT setting degrades when the X-ray dose becomes too low. When the dosage is decreased by more than 25\% compared to normal dosage, 
the manifestation of noise textures in IR methods tends to peak at lower spatial frequencies, leading to a worsened capability of IR methods to differentiate non-uniform regions and a decline in spatial resolution for low-contrast structures \cite{mileto2018ct}.

Data-driven methods for denoising LDCT images exploit the representation power of over-parametric models to build efficient and high-quality denoisers. The noise in the LDCT images is usually simulated by the mixed Poisson Gaussian distribution~\cite{Kaur_MPGD_2018}, which may demonstrate non-uniformity across regions with different smoothness and contrast in images. Supervised deep neural network denoisers
are expected to map LDCT images to normal dose CT images. An example is a U-Net-based framework FBPConvNet~\cite{2016Deep} that refines FBP reconstructed X-ray CT images. This can be viewed as a post-processing step that operates in the image domain.
Another supervised trained approach, WavResNet, operates in the wavelet domain to learn the residual from ground truth images in order to preserve detailed textures and edges.

W{\"u}rfl et al.\ leverage convolutional neural  networks to perform denoising in the sinogram domain~\cite{wurfl2016deep} and similarly,  
He et al.\ use fully connected layers to simulate back-projection and convolutional layers for subsequent denoising, thus attaining the effect of the inverse Radon transform~\cite{iRadonMap20}. 

A large type of supervised methods for LDCT are based on \textit{unrolling}, where a parametric model is incorporated into a series of iterates of an iterative optimization method and such a iterated block is seen as an end-to-end reconstruction operation. The parameters are then learned with supervision.
Examples include unrolling the
alternating direction method of multipliers (ADMM) algorithm \cite{admm-net2016}, primal-dual algorithms \cite{adler2018primaldual},
the block coordinate descent algorithm~\cite{Chun&etal:19MICCAI}, 
and the gradient descent algorithm \cite{adler2017solving}. 

We briefly discuss three types of unsupervised methods: generative adversarial networks (GAN) for denoising LDCT images, deep image prior (DIP), and transform-learning based regularization. The GAN approach addresses the lack of abundant paired data of low-dose images and corresponding normal-dose images. The generator aims to denoise aliased images while discriminators are supposed to discriminate denoised images output by the generator and ground truth images~\cite{Wolterink17}. Variants of GAN in terms of using different loss functions such as Wasserstein-GAN~\cite{WGAN_CT} and least-square GAN~\cite{LSGAN_CT} and modifying GAN training pipeline (cycle-GAN~\cite{cycleGAN_CT}) have also been used in the LDCT setting.
In the DIP setting, the network takes a fixed input of random noise, and aims to generate the noise-free image through minimizing a 
MBIR loss function
with respect to the target CT image~\cite{DIP-CT:20}. The guidance on when to stop DIP optimization process to prevent overfitting the noise is highly empirical. 
For the transform learning based optimization for LDCT denoising, Yang et al. utilize transforms which are organized in multiple layers to encode sparse representation of training CT images 
and 
incorporate this model into
a penalized weighted least squares (PWLS) optimization problem for 
reconstructing LDCT images~{\color{black}\cite{yang2022multilayer}}. 



Supervised learning methods often lead to lower denoising error compared to unsupervised learning methods 
when the distribution of the training data and testing data are similar. However, the required large paired data for supervised training may not always be available for applications such as LDCT.
Several unsupervised learning methods have the advantage of requiring unpaired and typically small datasets, and they often exhibit better generalization capabilities across different datasets.
For combining the advantages of supervised methods and unsupervised methods,
Ye et al.~\cite{2021Unified} 
presented a unified supervised-unsupervised learning framework for LDCT image reconstruction that combined a supervised deep learning approach and unsupervised transform learning (ULTRA) approach for robust LDCT image reconstruction, which we hereinafter refer to as \textit{serial SUPER}.
In the serial SUPER framework, the denoising process alternates between a neural network-based denoising step and an optimization step with a cost function consisting of a data-fidelity term and a pre-learned transform regularization term.

\subsection{Contribution}
In this work, we propose a method to combine supervised trained denoisers and transform learning-based unsupervised denoisers in a parallel manner (hereinafter referred to as \textit{parallel SUPER}) for improved LDCT image reconstruction. In each parallel SUPER block, an adaptive sparsifying-transform-based unsupervised reconstructor complements neural network based reconstructors. We show that with appropriate weights assigned to each reconstructor component, the parallel SUPER method provides better reconstructions than either component alone as well as the recent serial SUPER.
We extend our recent preliminary study \cite{2022Combining} in several aspects. First, we introduce a high-frequency filtering penalty into the loss function of the supervised module to improve streak artifacts and blurred details recovery in the reconstructed images. Meanwhile, we include a momentum term in the optimization formulation of the unsupervised reconstructor to ensure that the output from this pre-trained reconstructor continues to contribute new information as parallel SUPER blocks pile up. These additional regularization terms stabilize the training process and serve as empirical guidance for good convergence.
Secondly, we present more combinations of supervised and unsupervised methods in the proposed parallel SUPER framework and benchmark the parallel SUPER method with the deep boosting method \cite{Deepboosting2018}. We show that the parallel SUPER framework outperforms the deep boosting method in terms of  error evaluation criteria like RMSE, SSIM, and SNR, and better visual quality of reconstructed LDCT images.
Thirdly, we provide a quantitative method to determine the respective weight allocated to the output from the supervised trained neural network and the output from sparsifying transform-based penalized weighted least squares optimizers. 

\subsection{Structure of the Paper}
We organize this paper as follows.
In Section \uppercase\expandafter{\romannumeral2}, we describe the methodology and formulations of parallel SUPER. 
In Section \uppercase\expandafter{\romannumeral3}, we explain the algorithmic details for parallel SUPER.
In Section \uppercase\expandafter{\romannumeral4}, we present experimental results of the proposed method, compare results with several baseline image reconstruction methods, and study the quantitative properties of parallel SUPER, before concluding in Section \uppercase\expandafter{\romannumeral5}.

\section{Methodology of Parallel SUPER}
\label{sec::parallel_SUPER}
In this section we present the parallel SUPER formulation and explain how it is applied during inference time. We emphasize that the parallel SUPER approach could be used with different imaging modalities such as magnetic resonance imaging~\cite{SODUR_2022} and positron emission tomography~\cite{DIP-CT:20}. We give interpretations of our formulations, and discuss some specific examples of Parallel SUPER models for the low-dose CT application.


	


\subsection{Parallel SUPER Model}
\begin{figure*}[h]
	\centering
		\includegraphics[height=3cm]{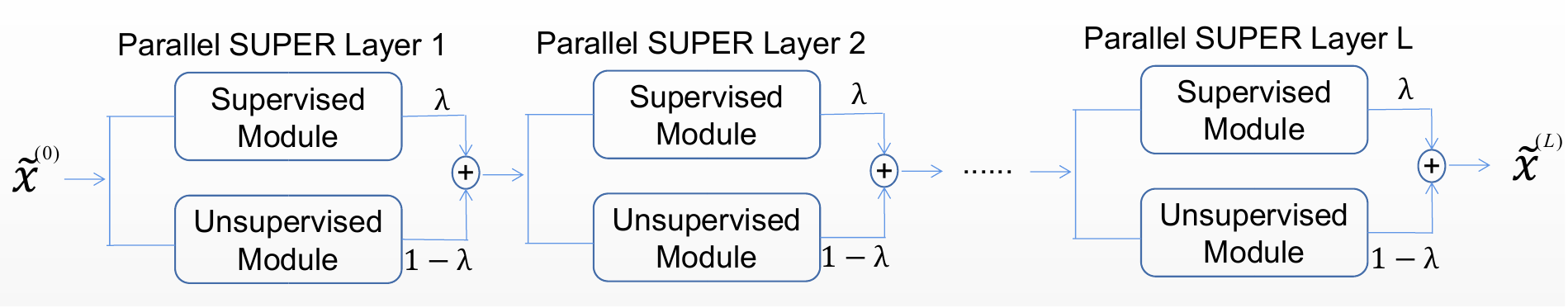}

	\vspace{-0.1in}
	\caption{Overview of the proposed Parallel SUPER Framework.}
	\label{fig::parallel_SUPER_paradigm}
\end{figure*}
The main idea of the Parallel SUPER model is to aggregate the supervised module and unsupervised module by a linear combination. Figure \ref{fig::parallel_SUPER_paradigm} shows the pipeline of our proposed Parallel SUPER method. This pipeline consists of multiple parallel SUPER blocks, each of which includes a Supervised module and an Unsupervised module. 


At the $l$th parallel SUPER block, the iterate is updated by the following relation:
\begin{equation}
\label{eqn::linear_combination}
  \widetilde{\mathbf{x}}^{(l)}_{n} =   {\lambda}^{(l)} G_{\theta^{(l)}}(\widetilde{\mathbf{x}}^{(l-1)}_{n})  + (1 - {\lambda}^{(l)}) \widehat{\mathbf{x}}^{(l)}_{n}(\mathbf{y}_{n}),
\end{equation}
where $n$ indicates the training sample index, 
$\widetilde{\mathbf{x}}^{(l)}_{n} \in \R^{N_p}$ is the output of the $l$-th parallel SUPER block,
${\y_{n}}\in \R^{N_d}$ is the noisy sinogram of {the} $n$-th image, 
$G_{\theta^{(l)}}(\widetilde{\mathbf{x}}^{(l-1)}_{n})$ is the output of the supervised module in the $l$-th parallel SUPER block, 
$\widehat{\mathbf{x}}^{(l)}_{n}$ is the output of the unsupervised module in the $l$-th parallel SUPER block,
and ${\lambda}^{(l)}$ is the weight parameter assigned to the supervised module in the $l$-th parallel SUPER block.


Define $\mathcal{M}(\widetilde{\x}^{(l-1)}_{n},\y_{n}; \Gamma)$ as an approximate 
(unsupervised) model-based iterative reconstruction (MBIR)  solver, which takes initial solution $\widetilde{\x}^{(l-1)}_{n}$, noisy sinogram data 
$\y_{n}$ and hyperparameter settings $\Gamma$ as the inputs and runs for a finite number of iterations to optimize the following problem:
\begin{align}
\label{eqn::unsupervised_obj}
\arg\min_{{\x_n\geq 0}} \underbrace{\frac{1}{2}\|\y_n-\mathbf{A}\x_n\|^2_{\mathbf{W}}}_{:= L(\mathbf{A}\x_n,\y_n)}   + \beta \mathcal{R}(\x_n) + \mu \| \x_n - 	\widetilde{\x}_n^{(l-1)}\|_2^2,
\end{align}
where $\mathbf{W} = \mathrm{diag}\{w_i\} \in \R^{N_d\times N_d}$ is a diagonal weighting matrix with diagonal elements $w_i$ set as the estimated inverse variance of $y_i$,  $\mathbf{A} \in \R^{N_d\times N_p}$ is the system matrix of the CT scan. In the MBIR problem \eqref{eqn::unsupervised_obj}, $L(\mathbf{A}\x_n,\y_n)$ is the data-fidelity term, penalty $\mathcal{R}(\x_n)$ is a fixed or pre-trained regularizer, and the parameter $\beta > 0$ controls the noise and resolution trade-off. 
We add an $\ell_2$ penalty term to make the iterative reconstruction method depend on the output from the previous Parallel SUPER block $\widetilde{\x}_n^{(l-1)}$, 
where $\mu \geq 0$ is a hyperparameter controlling the magnitude of this momentum-like regularization term. Our numerical experiments show this $\ell_2$ regularization term stabilizes the iterations in the parallel SUPER blocks.

We aggregate the supervised module and unsupervised module for the parallel SUPER method as the following:

\begin{algorithm}[t]\begin{footnotesize}
		\caption{Parallel SUPER Training Alogorithm}
		\label{alg::main}
		{\bf Input:} 
		$N$ pairs of reconstructed low-dose images and corresponding regular-dose reference images $\{(\widetilde{\x}^{(0)}_n, \x^{\star}_n)\}_{n=1}^{N}$, low-dose sinograms $\{\y_n\}$, weights $\mathbf{W}_n, \forall n$, number of parallel SUPER layers $L$, unsupervised module with corresponding hyperparameter settings $\mathcal{M}(\widetilde{\x}^{(l-1)}_{n},\y_{n}; \Gamma)$.
		
		{\bf Output:}  supervised module parameters {$\{{\param}^{(l)}\}_{l=1}^L$}, weights for supervised module $\{\lambda^{(l)}\}_{l=1}^L$.
		
		\begin{algorithmic}
			
			\FOR{$l=1,2, \dots , L$}

			\STATE\textbf{(1) Update $\widehat{\x}^{(l)}_n(\y_n)$:} using $\widetilde{\x}^{(l-1)}_n$ as initial image, using unsupervised module with corresponding hyperparameter settings $\mathcal{M}(\widetilde{\x}^{(l-1)}_{n},\y_{n}; \Gamma)$ to obtain each $\widehat{\x}^{(l)}_n(\y_n)$.

			\STATE\textbf{(2) Update $\bm{\theta}^{(l)}$:} with $N$ paired images $\{(\widetilde{\x}^{(l-1)}_n, \x^{*}_n)\}_{n=1}^{N}$, train the supervised model by solving optimization problem \eqref{eqn::supervised_obj} to update ${\param}^{(l)}$. 
			
			\STATE\textbf{(3) Update $\bm{\lambda}^{(l)}$:} with $N$ pairs of images $\{\widehat{\x}^{(l)}_n(\y_n), G_{\theta^{(l)}}(\widetilde{\mathbf{x}}^{(l-1)}_{n}), \x_{n}^{*}\}_{n=1}^{N}$ as reconstructions from two modules and ground truth, compute $\bm{\theta}^{(l)}$ according to \eqref{eqn::lambda_cond}.	
			
			\STATE\textbf{(4) Generate the output of $l$-th layer $\widetilde{\x}^{(l)}_n$:} with \eqref{eqn::linear_combination}, combine reconstructions from two modules to output to next parallel SUPER block.

			\ENDFOR\\
		\end{algorithmic}
	\end{footnotesize}
\end{algorithm}
\subsubsection{Parallel SUPER Learning Formulation}
We employ sequential training for the parallel SUPER model.
In the $l$-th layer, the parallel SUPER block is formulated as:
\begin{align}
	&\min_{{\lambda}^{(l)}}\sum_{n=1}^{N} \| G_{\param^{(l)}}(\widetilde{\mathbf{x}}^{(l-1)}_{n}) {\lambda}^{(l)} +\widehat{\mathbf{x}}^{(l)}_{n}(\mathbf{y}_{n})(1 - {\lambda}^{(l)} )  -  \mathbf{x}^{*}_{n}  \|_2^2 \nonumber  \\
	& \mathrm{s.t.} 
	\begin{cases}	
	\widehat{\x}^{(l)}_{n}(\y_{n}) = \mathcal{M}(\widetilde{\x}^{(l-1)}_{n},\y_{n}; \Gamma), \\
	 	\param^{(l)} = \mathrm{arg} \min_{\param^{(l)}} \sum_{n=1}^{N}\| G_{\param^{(l)}}(\widetilde{\x}^{(l-1)}_n) -  \x^{\star}_n  \|_2^2   + \alpha \| F(G_{\param^{(l)}}(\widetilde{\x}^{(l-1)}_n)) -  F(\x^{\star}_n ) \|_2^2\}, \\
 	\lambda_{\textrm{lb}} \le \lambda^{(l)} \le \lambda_{\textrm{ub}}
	\end{cases}
	\tag{P0}
	\label{eqn::parallel_super_target}
\end{align}
A full pipeline of parallel SUPER entails several blocks, and we optimize the parameters in each block in cascading order. Inside the optimization process of each parallel SUPER block, we solve problem \eqref{eqn::parallel_super_target} by respectively handling parameter update in the supervised module, computing an approximate solution to the optimization problem in the unsupervised module, and determining the combination parameter $\lambda^{(l)}$. 
The parallel SUPER learning algorithm based on \eqref{eqn::parallel_super_target} is illustrated in Algorithm \ref{alg::main}.

With respect to the supervised module, we set the loss function during training to be the root-mean-squared error (RMSE) to enforce alignment between the refined images and ground truth images. In the $l$-th parallel SUPER block, the optimization problem for training the neural network is:
\begin{equation}
    \mathop{\min}_{\param^{(l)}} \sum_{n=1}^{N}   \big\{\| G_{\param^{(l)}}(\widetilde{\x}^{(l-1)}_n) -  \x^{\star}_n  \|_2^2 + \alpha \| F(G_{\param^{(l)}}(\widetilde{\x}^{(l-1)}_n)) -  F(\x^{\star}_n ) \|_2^2 \big\},  \label{eqn::supervised_obj}
\end{equation}
where $G_{\param^{(l)}}(\cdot)$ denotes the neural network in the $l$-th parallel SUPER block with parameters $\param^{(l)}$, $\widetilde{\x}_n^{(l-1)}$ is the $n$-th input image from {the} $(l-1)$-th block,  $\x^{\star}_n$ is the corresponding normal-dose (reference) image or the training label. For better preserving the texture in the image, we add a term to the loss function of the supervised module to enforce the alignment of high frequency components between training pairs, and $F(\cdot)$ is a high frequency filter and $\alpha$ is the corresponding parameter. In practice, the high frequency filter $F$ can be a 
high pass filter in the Fourier domain or a Laplacian of Gaussian filter.

After obtaining the output of the supervised module and unsupervised module of the $l$-th Parallel SUPER block, we can choose the optimal $\lambda ^{(l)}$ to combine the outputs from both modules by solving the following optimization problem:
\begin{align}
\label{eqn::lambda_cond}
	\min_{\lambda^{(l)}}\sum_{n=1}^{N} \, &\| G_{\theta^{(l)}}(\widetilde{\mathbf{x}}^{(l-1)}_{n}) {\lambda}^{(l)} + \widehat{\mathbf{x}}^{(l)}_n(\mathbf{y}_{n})(1 - {\lambda}^{(l)} )  -  \mathbf{x}^{*}_{n}  \|_2^2 \nonumber  \\
	\text{ s.t. } &\lambda_{\textrm{lb}}\leq {\lambda}^{(l)} \leq \lambda_{\textrm{ub}},
\end{align}
where $\lambda_{\textrm{lb}}$ and $\lambda_{\textrm{ub}}$ indicate the low boundary and upper boundary of $\lambda$ and we restrict the value of the combination parameter in each parallel SUPER block between $\lambda_{\textrm{lb}}$ and $\lambda_{\textrm{ub}}$ to ensure both the supervised module and the unsupervised module contribute to the reconstruction non-trivially.
Solving \eqref{eqn::lambda_cond}, we get 
\begin{equation}
\lambda^{(l)} = \frac{\sum_{n=1}^{N}\big( \widehat{\mathbf{x}}^{(l)}_{n}(\mathbf{y}_{n})  -  G_{\theta^{(l)}}(\widetilde{\mathbf{x}}^{(l-1)}_{n})\big)^\top (\widehat{\mathbf{x}}^{(l)}_{n}(\mathbf{y}_{n}) - \mathbf{x}^{*}_{n})}{\sum_{n=1}^{N}\| G_{\theta^{(l)}}(\widetilde{\mathbf{x}}^{(l-1)}_{n}) - \widehat{\mathbf{x}}^{(l)}_{n}(\mathbf{y}_{n})\|_2^2}.
\end{equation}
We see that the weight assigned to the supervised module resembles the cosine similarity measurement between two differences: the discrepancy between supervised and unsupervised reconstruction in the $l$-th layer, and the discrepancy between the unsupervised reconstruction and the ground truth label. If the above $\lambda^{(l)} < \lambda_{\textrm{lb}}$, then we manually set $\lambda^{(l)} = \lambda_{\textrm{lb}}$. If the above $\lambda^{(l)} > \lambda_{\textrm{ub}}$, similarly we set $\lambda^{(l)} = \lambda_{\textrm{ub}}$. This solves \eqref{eqn::lambda_cond}.
With a determined $\lambda^{(l)}$, we obtain the output of the $l$-th Parallel SUPER block.

\subsubsection{Parallel SUPER Reconstruction for Testing}
With fixed supervised network parameters $\param^{(l)}$ and the combination parameter $\lambda^{(l)}$ in each parallel SUPER block in the pipeline, we can pass the testing data through the pipeline, obtain the output of each block which combines the output of supervised module and reconstructed image of unsupervised module, and take the output from the last block as the final testing result.

\section{Implementation Examples of Parallel SUPER}
This section introduces the algorithms to implement the parallel SUPER framework in Section \ref{sec::parallel_SUPER}. We give details about the supervised module FBPConvNet and unsupervised modules PWLS-ULREA and Plug-and-Play ADMM (PnP-ADMM) used in our experiments, and describe how they fit into the parallel SUPER framework.

The supervised module involves hierarchical neural network weights to effectively remove noise and artifacts and the unsupervised module can substantially optimize each image by leveraging a variety of physical and image attributes as priors. Our proposed method can combine their advantages and thus outperform each individual component. This framework has the flexibility to let users exploit any supervised method and unsupervised method 
for reconstruction.

\subsection{Composition Examples of Parallel SUPER Blocks}
We present examples to compose a parallel SUPER block by choosing different neural networks $G(\cdot)$ in the supervised module and different regularizers $R( \cdot )$ in the unsupervised module. For 
neural networks in the supervised module, we use the FBPConvNet (FCN). For 
regularizers in the unsupervised module, we experiment with the PWLS-ULTRA method and the PnP-ADMM method. We hereinafter refer to the resulting parallel SUPER realizations as PS-FCN-ULTRA and PS-FCN-PnP, respectively.

\subsubsection{Supervised Module}
FBPConvNet is a U-Net based image-domain denoising architecture originally designed
for sparse-view CT. We apply FBPConvNet here to the low-dose CT setting where it learns to map low-dose FBP images as network input to corresponding high-quality reference images. Traditional U-Net uses a dyadic scale decomposition based on max pooling. Similar to U-Net, FBPConvNet adopts multichannel filters to increase the capacity of the network.

\subsubsection{Unsupervised Module}
For the unsupervised module, we choose the PWLS-ULTRA method~\cite{2018SPULTRA} and the PnP-ADMM~\cite{plug_and_play} with BM3D denoiser~\cite{BM3D}.

The PWLS-ULTRA method reconstructs an image $\x$ from noisy sinogram data  $\y$ (measurements) with a union of pre-learned transforms $\{ \boldsymbol{\Omega}_k \}_{k=1}^{K}$. The image reconstruction is done through the following nonconvex optimization problem:
\begin{equation}
        \arg \min_{\x} \bigg\{ \frac{1}{2}\|\y-\mathbf{A}\x\|^2_{\mathbf{W}} +  \min_{\mathcal{C}_k, \mathbf{z}_j} \beta  \sum_{k=1}^{K} \sum_{j \in \{\mathcal{C}_k\}_{k=1}^K} \bigg( \|\boldsymbol{\Omega}_k \mathbf{P}_j \x - \mathbf{z}_j \|_2^2 + \gamma^2 \|\mathbf{z}_j\|_0 \bigg) +\mu \| \x - 	\widetilde{\x}^{(l-1)}\|_2^2 \bigg\}.
\end{equation}

$\widehat{\x}^{(l)}(\y)$ denotes the reconstructed image by the unsupervised solver over several iterations in the $l$-th layer, the operator $\mathbf{P}_j \in \R^{l \times N_p}$ above extracts the $j$-th patch of $l$ voxels of image $\x$ as $\mathbf{P}_j \x$, $\mathbf{z}_j$ is the corresponding sparse encoding of the image patch under a matched transform, and $\mathcal{C}_k$ denotes the indices of patches grouped into the $k$-th cluster with transform $\boldsymbol{\Omega}_k$. Minimization over $\mathcal{C}_k$ indicates the computation of the cluster assignment of each patch.
The regularizer $\mathcal{R}$ above includes an encoding error term and an $\ell_0$ sparsity penalty (with weight $\gamma^2$) counting the number of non-zero entries and $\beta$ is the regularization parameter. 
We apply the alternating minimization method from~\cite{zheng2018} (with inner iterations when updating $\x$) on the above optimization problem. 
The sparse encoding and clustering are computed simultaneously~\cite{zheng2018}.
In the parallel SUPER implementation, using different initialization in each parallel SUPER block benefits optimizing the nonconvex problem \eqref{eqn::unsupervised_obj}.

The PnP-ADMM 
involves the use of a standard off-the-shelf denoising algorithm as a replacement for the proximal operator inside an optimization algorithm for model-based image reconstruction.
We split the variable $\x$ into $\mathbf{v}$ in \eqref{eqn::unsupervised_obj}, and use ADMM to solve the resulting optimization problem:
\begin{equation}
\label{eqn::split_unsupervised_obj}
	(\hat{\x}, \hat{\mathbf{v}}) = \mathrm{arg} \min_{{\x, \mathbf{v}}, {\x = \mathbf{v}}} {\frac{1}{2}\|\y-\mathbf{Ax}\|^2_{\mathbf{W}}}  + \beta \mathcal{R}(\mathbf{v}) + \mu \| \x - 	\widetilde{\x}\|_2^2. 
\end{equation}

\subsection{More Details about Unsupervised Modules}
\subsubsection{Learning a Union of Sparsifying Transforms}
We pre-learn a union of transforms $\{\omg_k\}_{k=1}^K$ to effectively group and sparsify a training set of image patches by solving:

\begin{equation}
\min_{\{\omg_k,\,\Z_i,\,\mathcal{C}_k \}} \sum_{k=1}^{K} \sum_{i\in \mathcal{C}_k} \left\{ \|\omg_k \X_i-\Z_i \|_2^2 +\eta^2\|\Z_i\|_0 \right\} +\sum_{k=1}^{K} \lambda_k Q(\omg_k),\quad \textup{s.t.} \quad \{\mathcal{C}_k \} \in \mathcal{{G}},
\label{eq:train_ultra}
\end{equation}
where $\X_i\in\mathbb{R}^m$ denotes the $i$-th vectorized (overlapping) image patch extracted from training images, $\Z_i\in\mathbb{R}^m$ is the corresponding transform-domain sparse approximation, parameter $K$ denotes the number of clusters, $\mathcal{C}_k$ denotes the indices of all the patches matched to the $k$-th transform, and the set $\mathcal{G}$ is composed of of all possible partitions of $\{1,2,\dots,N'\}$ into $K$ disjoint subsets, with $N'$ denoting the total number of training patches.
Each transform $\omg_k$ has its regularizer term $Q(\omg_k)=\|\omg_k\|_\mathrm{F}^2-\log|\det\omg_k| \textrm{ for } 1\le k \le K$, which controls properties of the transform (regularizes the scaling and condition number of the transform), and prevents trivial solutions (e.g., matrices with zero or repeated rows).
We set the weights $\lambda_k=\lambda_0 \sum_{i \in \mathcal{C}_k} \|\X_{i}\|^2_2$, where $\lambda_0$ is a constant \cite{zheng2018}.
We adopt an alternating algorithm to solve optimization problem \eqref{eq:train_ultra} that alternates between a \textit{transform update step} (solving for $\{\omg_k\}$) and a \textit{sparse coding and clustering step } (solving for $\{\Z_i,\,C_k\}$), with closed-form solutions adopted in each step.
As a patch-based unsupervised learning method, ULTRA typically only needs a few regular-dose training images to learn rich features.

\subsubsection{Plug-and-Play ADMM Method}
The PnP-ADMM is a modification of the conventional ADMM. We can iteratively update the variables using ADMM to solve \eqref{eqn::split_unsupervised_obj} as follows~\cite{chan2016plug}:
\begin{align}
\label{eqn::admmx} 
\x^{(k+1)} &= \mathrm{arg}\min_{\x} \frac{1}{2}\|\y-\mathbf{A}\x\|^2_{\mathbf{W}}  + \mu \| \x - \widetilde{\x}\|_2^2 + \frac{\rho_k}{2}\|\x - (\mathbf{v}^{(k)} - \mathbf{u}^{(k)}) \|_2^2 \\
\label{eqn::admmv}
\mathbf{v}^{(k+1)}&= \mathrm{arg}\min_{{\mathbf{v}}} \mathcal{R}(\mathbf{v}) + \frac{1}{2\frac{\beta}{\rho_k}}\|	\mathbf{v} - (\x^{(k+1)}+\mathbf{u}^{(k)}) \|_2^2\\
\label{eqn::admmu}
\mathbf{u}^{(k+1)} &= \mathbf{u}^{(k)} + ( \x^{(k+1)} - \mathbf{v}^{(k + 1)} )\\
\rho_{k+1} &= \gamma_k\rho_k, \textrm{ where } \gamma_k\ge 1. \nonumber
\end{align}

The optimization problem in \eqref{eqn::admmv} can be abstracted into an off-the-shelf image denoising step (denoiser), which we denote as
$	\mathbf{v}^{(k+1)} = D_{\sigma_k} (\x^{(k+1)}+\mathbf{u}^{(k)})$ 
where $\sigma_k=\sqrt{\frac{\beta}{\rho_k}}$. Notice that a pre-specified denoiser can implicitly define a regularizer $\mathcal{R}$. PnP-ADMM has convergence guarantee with the assumption that the used denoiser is bounded in the sense that $\|D_{\sigma}(\x)-\x\|^2/n \le \sigma^2 C$ for some constant $C>0$ ($\x\in\R^n$).
 One can use any denoiser in this PnP-ADMM scheme and we have adopted the BM3D method in this study.

\section{Experiments and Discussions}

In this section, we first introduce a deep boosting 
method
as our baseline. Then we introduce the experimental setup and the details of our proposed methods and other methods used for comparison. 
We show both numerical results and visual results for comparison.
Finally, we discuss the effect of the regularization parameters $\mu$ and $\beta$ and show the convergence behavior of our proposed method.
\subsection{Baseline: Deep Boosting Model}
One baseline with which we benchmark our proposed parallel SUPER method is the deep boosting framework. \textit{Boosting} is an established machine learning technique that leverages an ensemble of weak learners to build a strong model in a sequentially trained manner.
Essentially, each step of boosting aims to extract the residual signal left from the previous weak learners, and expects that the newly added base learner can pick up the leftover signal. 
Classical boosting base learner usually involves adaptive basis functions such as regression trees or neural networks. 
In the regression setting, gradient boosting can be interpreted as  gradient descent in the functional space, as each added weak learner is a step towards the optimal function that minimizes the target objective.

In \cite{2015Boosting}, the Strengthen-Operate-Subtract (SOS) boosting framework is proposed to amplify signal-to-noise ratio (SNR) through a specific recursion format. 
The core iteration design is through $\widehat{\x}^{k+1} = f(\x_0 + \widehat{\x}^k) - \widehat{\x}^k$, where $k$ is the boosting layer, $\x_0$ is the noisy image and $\widehat{\x}^k$ is the output of the $k$-th boosting layer.
The denoising target of deep boosting method in each iteration step
is the strengthened image $\x_0 + \widehat{\x}^k$, instead of the residual image $\x_{\textrm{True}} - \widehat{\x}^k$.
In the first term, the signal is strengthened by adding the previous denoised image $\widehat{\x}^k$ to the noisy initialization $\x_0$ given measurements, and $f$ carries out the denoising protocol on the strengthened compound signal.
The subtraction of the denoised image $\widehat{\x}^k$ from the output of the denoising operation $f$ gives the new denoised image $\widehat{\x}^{k+1}$. The motivation for the SOS protocol is the observation that $\textrm{SNR}(\x_0 + \widehat{\x}) > \textrm{SNR}(\x_0)$. We expect $\|\x_0-\x_{\textrm{True}}\| \gg \|\widehat{\x}-\x_{\textrm{True}}\|$ assuming the effectiveness of the denoising procedure, therefore this SNR relationship can be seen by $\textrm{SNR}(\x_0+\widehat{\x}) = \frac{\|2\x_{\textrm{True}}\|}{\|(\x_0-\x_{\textrm{True}}) + (\widehat{\x}-\x_{\textrm{True}})\|} \ge \frac{2\|\x_{\textrm{True}}\|}{(1+\delta)\|\x_0-\x_{\textrm{True}}\|} = \frac{2}{1+\delta} \textrm{SNR}(\x_0)$ for some $\delta\ll 1$. 

\paragraph{Deep Boosting Learning Formulation} \cite{Deepboosting2018} suggests a boosting framework based on the SOS design where a series of CNNs is used to denoise images. 
With $\{G_n\}$ as a series of CNNs, the   iteration scheme is suggested as:
\begin{align*}
    \x_{1} &= G_{1}(\x_0), \\
    \x_{n} &= G_{n}\big( \x_0 + \x_{n-1} \big), \textrm{ for } n>1,
\end{align*}
and the loss function for training is taken as $L(\widehat{\x}_n, \x_{\textrm{True}}) = \|\widehat{\x}_n - \x_{\textrm{True}}\|_2^2$. We see that the denoisers $G_n$ also take as input the strengthened signal $\x_0+\x_{n-1}$. The motivation for doing so is to enlarge the SNR as stated in the previous paragraph.

\paragraph{Shortcomings of the boosting framework} In the gradient boosting scheme, when the iteration goes deeper and residual signals become weaker, the scaling issue becomes more dominant despite the strong expressiveness of the neural network. In other words, the signal becomes so small in scale so that base-type regressors can no longer capture the remaining signal in the residual and thus adding more regressors to the ensemble does not help any further. The deep boosting learning formulation partially mitigates this issue by letting regressors to denoise strengthened signals instead of learning residuals. Nonetheless, repeatedly applying the denoising function is unlikely to lead to the continual decay of noise magnitude in the iteration sequence, as indicated in Figure \ref{fig::RMSEevolution}.

In our work, we attempt a combination of supervised and unsupervised methods to extend the capability of the extant methods to reduce the denoising error and extract finer patterns from underlying signals. We show that the final reconstruction quality from parallel SUPER outperforms the deep boosting baseline.

\subsection{Experiment Setup}

\subsubsection{Data and Imaging System}
In our experiments, we use the Mayo Clinics dataset established for ``the 2016 NIH-AAPM-Mayo Clinic Low Dose CT Grand Challenge'' \cite{2016TU}. We choose 520 images from 6 of 10 patients in the dataset, among which 500 slices are used for training and 20 slices are used for validation. We randomly selected 20 images from the remaining 4 patients for testing. We project the regular dose CT images $\x^{\star}$ to sinograms $\y$ and add Poisson and additive Gaussian noise to them as follows: 
\begin{equation*}
	\centering
	y_{i}=- \log \left( I_0^{-1}\max\big(\textup{Poisson}\{ I_0 e^{-[\mathbf{Ax}^\star]_i}\} + \mathcal{N}\{0,\,\sigma^2\},\varepsilon \big) \right),
\end{equation*} 
where the original number of incident photons per ray is $I_0=10^4$, the Gaussian noise variance is $\sigma^2=25$, and $\varepsilon$ is a small positive number to avoid negative measurement data when taking the logarithm~\cite{2018SPULTRA}.

We use {the Michigan Image Reconstruction Toolbox to construct} fan-beam CT geometry with 736 detectors~$\times$~1152 regularly spaced projection views, and a no-scatter mono-energetic source. 
{The width of each detector column is 1.2858~mm, the source to detector distance is 1085.6~mm, and the source to rotation center distance is 595~mm. We reconstruct}
images of size $512\times 512$ with the pixel size being 0.69~mm $\times$ 0.69~mm.

\subsubsection{Baselines} 
We compare the proposed parallel SUPER model with the unsupervised method (PWLS-EP), the standalone unsupervised module (PWLS-ULTRA), the standalone unsupervised module (PnP-ADMM), the standalone supervised module (FBPConvNet), the serial SUPER model and the deep boosting model.
PWLS-EP is a penalized weighted-least squares reconstruction method with edge-preserving hyperbola regularization. For the unsupervised method (PWLS-EP), we set the scale parameter of potential function $\delta = 20$ and regularization parameter $\beta = 2 ^ {15}$ and run 100 iterations to obtain convergent results.
For the standalone unsupervised module (PWLS-ULTRA), we have trained a union of $5$ sparsifying transforms using $12$ slices of regular-dose CT images with parameters 
 $\beta=5\times 10^{3}$ and $\gamma=20$.  In reconstruction, we set the parameters $\beta= 10^{4}$ and $\gamma=25$, and run $1000$ alternating steps with 5 inner iterations in each step to attain reported reconstruction performance.
For the standalone unsupervised module (PnP-ADMM),  we set the parameters $\rho_0=10^{6}$, $\gamma_k=1$ and $\beta = 25$ for the BM3D denoiser, and run $1000$ alternating steps to achieve convergence.
When training the standalone supervised module (FBPConvNet), we run 100 epochs to sufficiently learn the image features with low overfitting risks.
For the serial SUPER model, we run $4$ epochs of FBPConvNet training for its supervised modules, and we use the pre-learned union of $5$ sparsifying transforms and set the parameters  $\beta=5\times 10^{3}$, $\gamma=20$ and $\mu = 5 \times 10^{5}$ to reconstruct images with $20$ alternating steps and $5$ inner iterations for the unsupervised module (PWLS-ULTRA). This specific serial SUPER model is named SS-FCN-ULTRA.
For the deep boosting method, we use FBPConvNet as the base denoiser in the boosting framework, which is referred to as Boosting (i.e.\ $G_n, \forall n$ is FBPConvNet).

\subsubsection{Parameter Settings}
In the parallel SUPER model, we employ FBPConvNet as the supervised module and either PWLS-ULTRA or the PnP-ADMM method as the unsupervised module. These two variations of the parallel SUPER model are named PS-FCN-ULTRA and PS-FCN-PnP, respectively.
It takes about 10 hours for training the model for 10 blocks in a GTX Titan GPU. 
For the cost function \eqref{eqn::supervised_obj} of the supervised method, we use the  Laplacian of Gaussian filter with standard deviation $0.5$ and kernel size $15\times 15$, and the parameter $\alpha = 10$.
During the training of the supervised method, we run $4$ epochs (kept small to reduce overfitting risks) with the stochastic gradient descent (SGD) optimizer for the FBPConvNet module in each parallel SUPER block. The training hyperparameters of FBPConvNet are set as follows: the learning rate decreases logarithmically from 0.001 to 0.0001; the batchsize is 1; and the momentum parameter is 0.99. The filters of networks across all blocks are initialized before training with i.i.d.\ random Gaussian entries with zero mean and variance 0.005. 
For the PWLS-ULTRA method,  We use the pre-learned union of $5$ sparsifying transforms in standalone PWLS-ULTRA method to reconstruct images with $5$ outer iterations and $5$ inner iterations of PWLS-ULTRA. We set the parameters  $\mu=5\times 10^{5}$, $\beta=5\times 10^{3}$ and $\gamma=20$ in reconstruction.
For the PnP ADMM method, we used the BM3D method as the denoiser and set the parameters $\mu=5\times 10^{5}$, $\rho_0=10^{6}$, $\gamma_k = 1$ and $\beta = 25$ for the BM3D denoiser.
PWLS-EP reconstruction is used as the initialization $\widetilde{\x}^{(0)}$ of the input of networks in the first layer.

\subsubsection{Evaluation Metrics}
We use root mean square error (RMSE), signal-to-noise ratio (SNR), and structural similarity index measure (SSIM) \cite{xu:12:ldx} to quantitatively evaluate the performance of reconstruction methods.
The RMSE in Hounsfield units (HU) is defined as $\textup{RMSE}=\sqrt{\sum_{j=1}^{N_p}(\hat{\x}_j-\x^*_j)^2/N_p}$, where $\x^*_j$ is the $j$th pixel of the reference regular-dose image $\x^*$, $\hat{\x}_j$ is the $j$th pixel of the reconstructed image $\hat{\x}$, and $N_p$ is the number of pixels.
The SNR in decibels (dB) is defined as $\textup{SNR}=10\log_{10} \frac{\|\x^*\|^2}{\|\hat{\x}-\x^*\|^2}$.

\subsection{Numerical Results and Comparisons}
\begin{table}[h]
	\centering
	\caption{Mean metrics of reconstructions for 20 test slices.}
	\scalebox{1.1}
    {\begin{tabular}{cccc}
		\toprule
		\textbf{Method} &
		\shortstack{\textbf{ RMSE (HU)}}&
		\textbf{ SNR (dB)}&
		\textbf{ SSIM} \\  \midrule
  
		PWLS-EP &41.4 &25.4 &0.673 \\ 
		PWLS-ULTRA &32.4 &27.8 &0.716 \\ 
		FBPConvNet & 29.2 & 28.2 & 0.688 \\
          PnP-ADMM & 35.4 & 27.1 & 0.69 \\
        SS-FCN-ULTRA & \BLUE{25.0} & 29.7 & 0.748 \\
        PS-FCN-ULTRA & \textbf{22.0} & \textbf{30.6} & \textbf{0.752} \\
        PS-FCN-PnP & 23 & 30.3 & 0.75 \\
       {Boosting} & 27.6 & 28.6 & 0.718 \\
		\bottomrule
	\end{tabular}}
	\label{tab:results}

\end{table}
\begin{figure}[h]
	\centering
		\includegraphics[height=5cm]{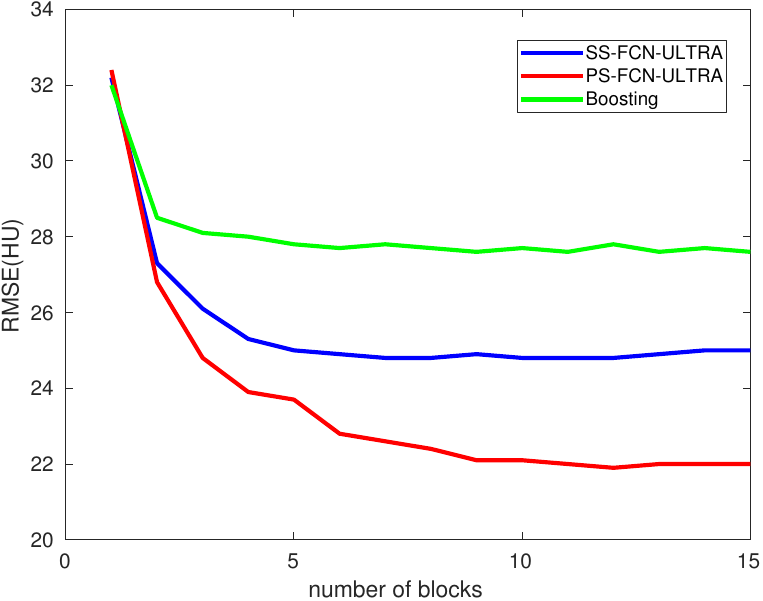}
	\caption{Mean RMSE evolution (over 20 test slices) of the SS-FCN-ULTRA, PS-FCN-ULTRA and Boosting  methods.}\label{fig::RMSEevolution}
\end{figure}

\begin{figure}[!t]
	\centering
	\begin{subfigure}{0.48\textwidth}
		\centering
		\includegraphics[width=\textwidth]{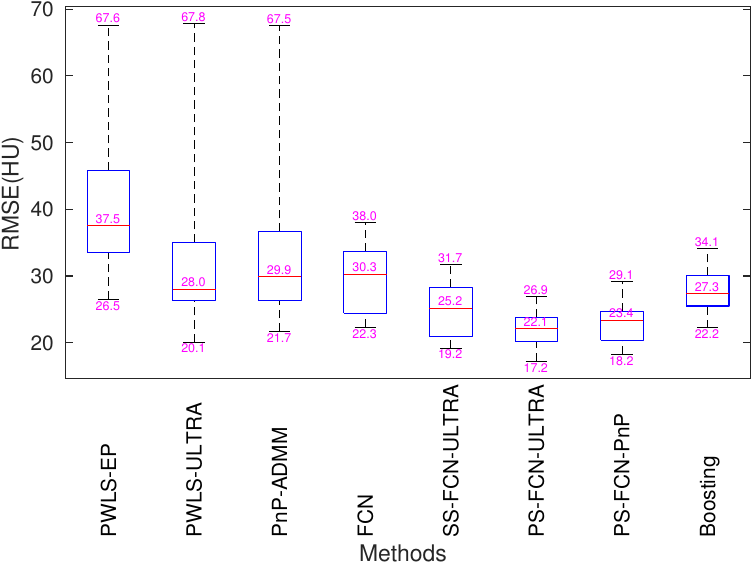}
		\caption{RMSE}
		\label{fig:rmsebox}
	\end{subfigure}
	\begin{subfigure}{0.49\textwidth}
		\centering
		\includegraphics[width=\textwidth]{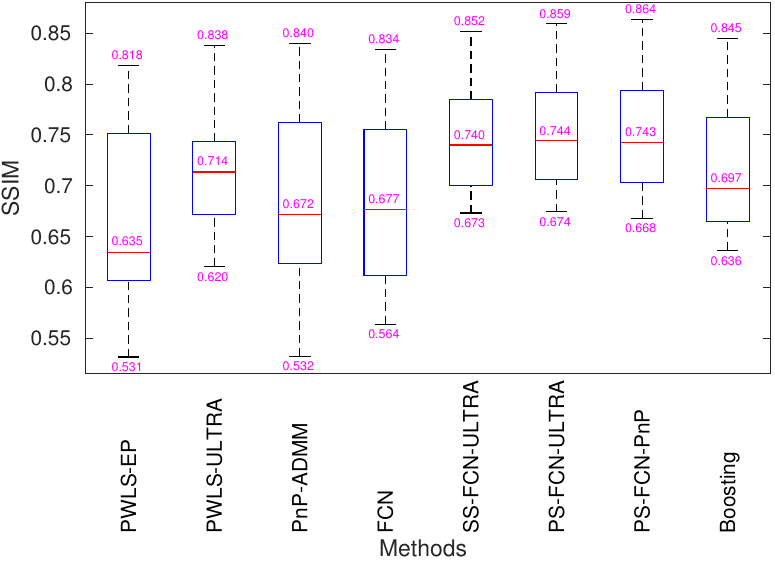}
		\caption{SSIM}
		\label{fig:ssimbox}
	\end{subfigure}
	\begin{subfigure}{0.49\textwidth}
		\centering
		\includegraphics[width=\textwidth]{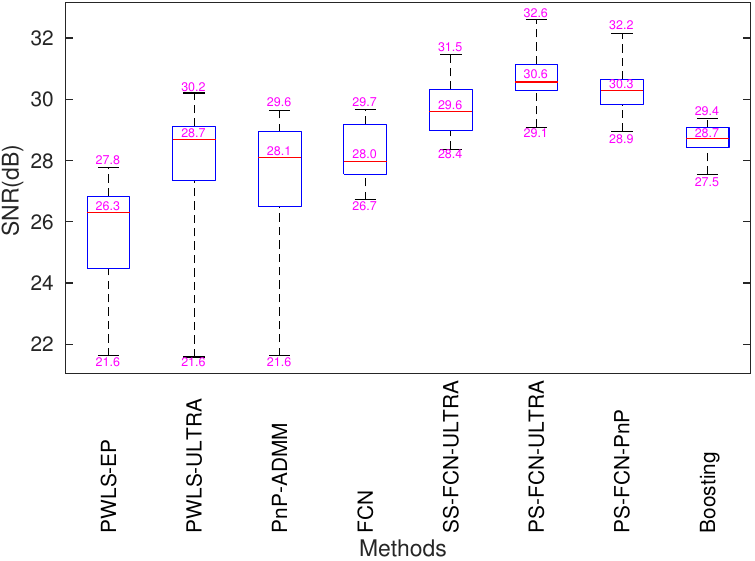}
		\caption{SNR}
		\label{fig:snrbox}
	\end{subfigure}
	\caption{RMSE, SSIM and SNR spread (shown using box plots) over 20 test cases using different methods. Each box plot for a method describes the statistics of RMSE, SSIM and SNR values over the 20 test slices: the central red line indicates the median; the bottom and top edges of the boxes indicate the 25th and 75th percentiles, respectively; and the whiskers represent the extreme values. We also marked the median values and the extreme values for each method in the box plot.  The proposed Parallel SUPER structures outperform both the standalone supervised module and  standalone unsupervised counterparts.
    They also outperform the boosting method and the Serial SUPER structure.}
\end{figure}

\begin{figure*}[!t]
	\centering  
  	\begin{tikzpicture}
        \begin{scope}
		[spy using outlines={rectangle,green,magnification=2,size=9mm, connect spies}]
		\node {\includegraphics[width=0.24\textwidth]{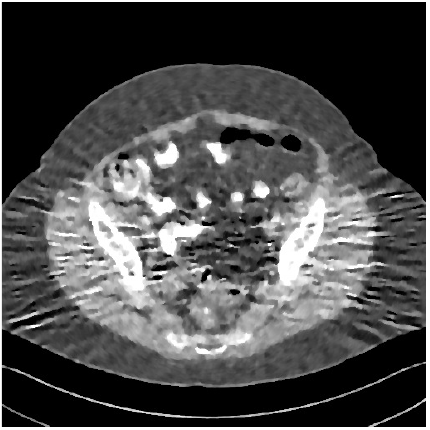} };
		\spy [green, draw, height = 0.77cm, width = 0.77cm, magnification = 2,
		connect spies] on (1.36,-0.5) in node [left] at (2,1.65);
  		\spy [green, draw, height = 0.77cm, width = 0.77cm, magnification = 2,
		connect spies] on (-0.57,-0.93) in node [left] at (-1.25,-1.65);
        \end{scope}
        \draw[red, >=stealth, ->] (2.02,2.02) -- (1.83,1.85);
        \draw[red, >=stealth, ->] (-1.93,-1.95) -- (-1.68,-1.88);
	\end{tikzpicture}
	\put(-80,6){ \color{white}{\bf \small{RMSE:46.6HU}}}
	\put(-115,105){ \color{white}{\bf \small{PWLS-EP}}} 	
 		\begin{tikzpicture}
   \begin{scope}
		[spy using outlines={rectangle,green,magnification=2,size=9mm, connect spies}]
		\node {\includegraphics[width=0.24\textwidth]{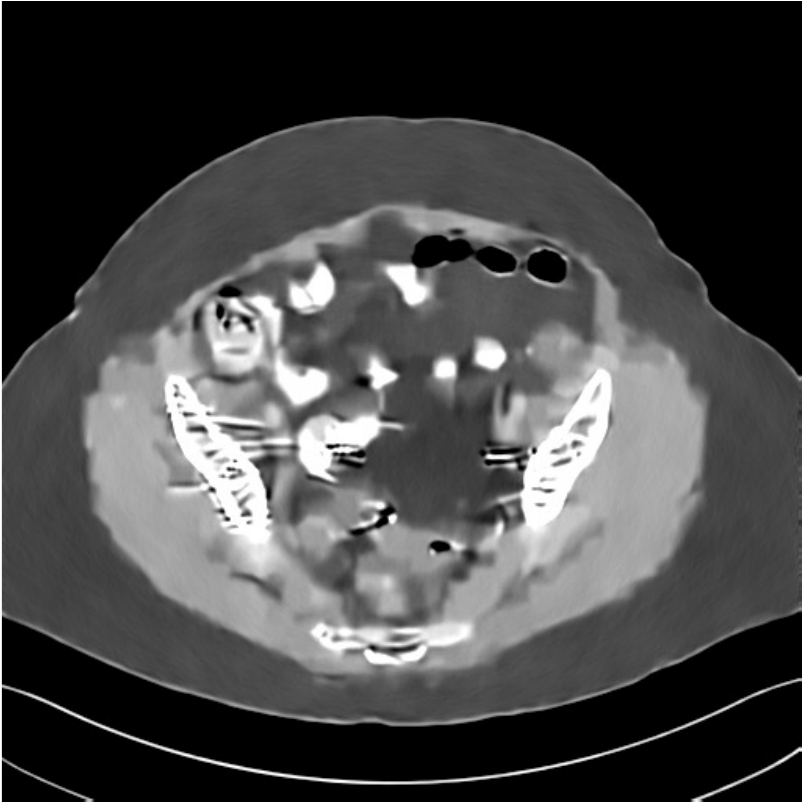} };
            \spy [green, draw, height = 0.77cm, width = 0.77cm, magnification = 2,
		connect spies] on (1.36,-0.5) in node [left] at (2,1.65);
  		\spy [green, draw, height = 0.77cm, width = 0.77cm, magnification = 2,
		connect spies] on (-0.57,-0.93) in node [left] at (-1.25,-1.65);
        \end{scope}
        \draw[red, >=stealth, ->] (2.02,2.02) -- (1.83,1.85);
        \draw[red, >=stealth, ->] (-1.93,-1.95) -- (-1.68,-1.88);
	\end{tikzpicture}
	\put(-80,6){ \color{white}{\bf \small{RMSE:34.6HU}}}
	\put(-115,105){ \color{white}{\bf \small{PWLS-ULTRA}}} 	
 	\begin{tikzpicture}
  \begin{scope}
		[spy using outlines={rectangle,green,magnification=2,size=9mm, connect spies}]
		\node {\includegraphics[width=0.24\textwidth]{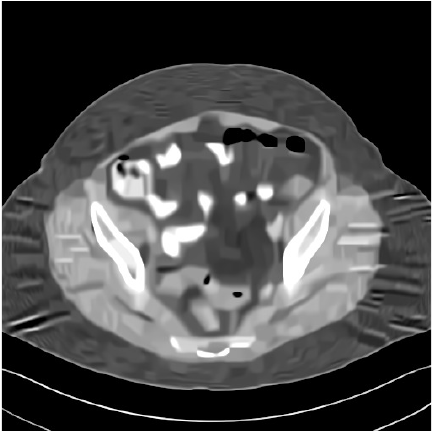} };
            \spy [green, draw, height = 0.77cm, width = 0.77cm, magnification = 2,
		connect spies] on (1.36,-0.5) in node [left] at (2,1.65);
  		\spy [green, draw, height = 0.77cm, width = 0.77cm, magnification = 2,
		connect spies] on (-0.57,-0.93) in node [left] at (-1.25,-1.65);
        \end{scope}
        \draw[red, >=stealth, ->] (2.02,2.02) -- (1.83,1.85);
        \draw[red, >=stealth, ->] (-1.93,-1.95) -- (-1.68,-1.88);
  \end{tikzpicture}
	\put(-80,6){ \color{white}{\bf \small{RMSE:35HU}}}
	\put(-115,105){ \color{white}{\bf \small{PnP-ADMM}}} 
	\begin{tikzpicture}
 \begin{scope}
		[spy using outlines={rectangle,green,magnification=2,size=9mm, connect spies}]
		\node {\includegraphics[width=0.24\textwidth]{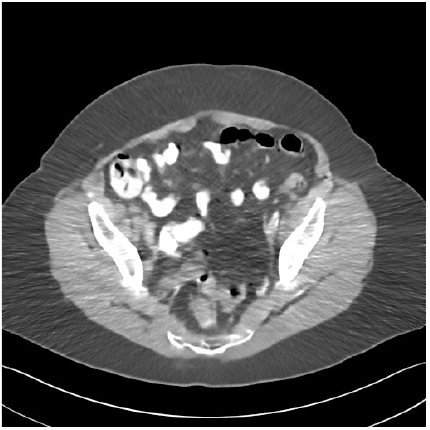} };
            \spy [green, draw, height = 0.77cm, width = 0.77cm, magnification = 2,
		connect spies] on (1.36,-0.5) in node [left] at (2,1.65);
  		\spy [green, draw, height = 0.77cm, width = 0.77cm, magnification = 2,
		connect spies] on (-0.57,-0.93) in node [left] at (-1.25,-1.65);
        \end{scope}
        \draw[red, >=stealth, ->] (2.02,2.02) -- (1.83,1.85);
        \draw[red, >=stealth, ->] (-1.93,-1.95) -- (-1.68,-1.88);
	\end{tikzpicture}
	\put(-80,6){ \color{white}{\bf \small{RMSE:32.6HU}}}
	\put(-115,105){ \color{white}{\bf \small{FBPConvNet}}} 	
 
	\begin{tikzpicture}
 \begin{scope}
		[spy using outlines={rectangle,green,magnification=2,size=9mm, connect spies}]
		\node {\includegraphics[width=0.24\textwidth]{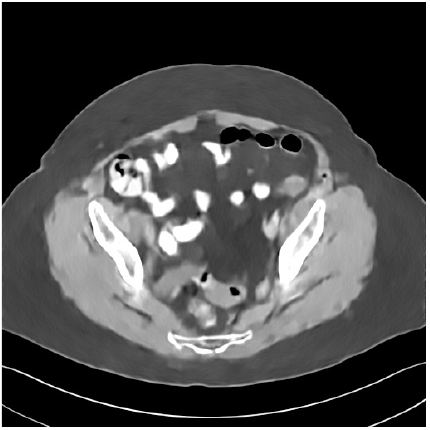}};
            \spy [green, draw, height = 0.77cm, width = 0.77cm, magnification = 2,
		connect spies] on (1.36,-0.5) in node [left] at (2,1.65);
  		\spy [green, draw, height = 0.77cm, width = 0.77cm, magnification = 2,
		connect spies] on (-0.57,-0.93) in node [left] at (-1.25,-1.65);
        \end{scope}
        \draw[red, >=stealth, ->] (2.02,2.02) -- (1.83,1.85);
        \draw[red, >=stealth, ->] (-1.93,-1.95) -- (-1.68,-1.88);
	\end{tikzpicture}
	\put(-80,6){ \color{white}{\bf \small{RMSE:26.5HU}}}
	\put(-115,105){ \color{white}{\bf \small{SS-FCN-ULTRA}}} 
    \begin{tikzpicture}
    \begin{scope}
		[spy using outlines={rectangle,green,magnification=2,size=9mm, connect spies}]
		\node {\includegraphics[width=0.24\textwidth]{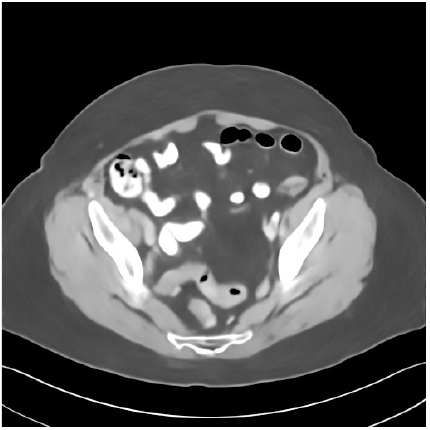}};
            \spy [green, draw, height = 0.77cm, width = 0.77cm, magnification = 2,
		connect spies] on (1.36,-0.5) in node [left] at (2,1.65);
  		\spy [green, draw, height = 0.77cm, width = 0.77cm, magnification = 2,
		connect spies] on (-0.57,-0.93) in node [left] at (-1.25,-1.65);
        \end{scope}
        \draw[red, >=stealth, ->] (2.02,2.02) -- (1.83,1.85);
        \draw[red, >=stealth, ->] (-1.93,-1.95) -- (-1.68,-1.88);
	\end{tikzpicture}
	\put(-80,6){ \color{red}{\bf \small{RMSE:22.1HU}}}
	\put(-115,105){ \color{red}{\bf \small{PS-FCN-ULTRA}}}
 	\begin{tikzpicture}
  \begin{scope}
		[spy using outlines={rectangle,green,magnification=2,size=9mm, connect spies}]
		\node {\includegraphics[width=0.24\textwidth]{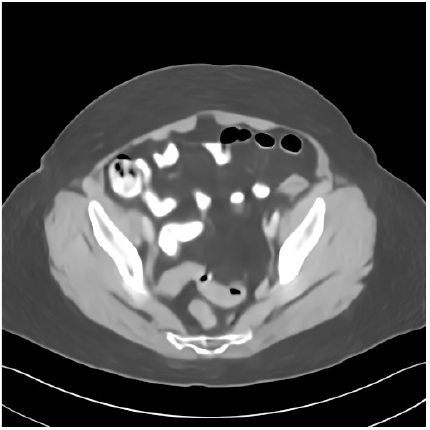} };
            \spy [green, draw, height = 0.77cm, width = 0.77cm, magnification = 2,
		connect spies] on (1.36,-0.5) in node [left] at (2,1.65);
  		\spy [green, draw, height = 0.77cm, width = 0.77cm, magnification = 2,
		connect spies] on (-0.57,-0.93) in node [left] at (-1.25,-1.65);
        \end{scope}
        \draw[red, >=stealth, ->] (2.02,2.02) -- (1.83,1.85);
        \draw[red, >=stealth, ->] (-1.93,-1.95) -- (-1.68,-1.88);
	\end{tikzpicture}
	\put(-80,6){ \color{red}{\bf \small{RMSE:23HU}}}
	\put(-115,105){ \color{red}{\bf \small{PS-FCN-PnP}}} 
    \begin{tikzpicture}
     \begin{scope}
		[spy using outlines={rectangle,green,magnification=2,size=9mm, connect spies}]
		\node {\includegraphics[width=0.24\textwidth]{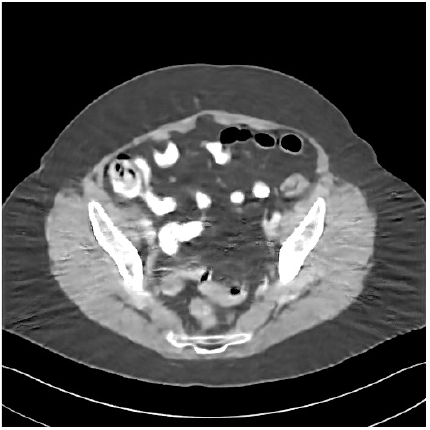} };
            \spy [green, draw, height = 0.77cm, width = 0.77cm, magnification = 2,
		connect spies] on (1.36,-0.5) in node [left] at (2,1.65);
  		\spy [green, draw, height = 0.77cm, width = 0.77cm, magnification = 2,
		connect spies] on (-0.57,-0.93) in node [left] at (-1.25,-1.65);
        \end{scope}
        \draw[red, >=stealth, ->] (2.02,2.02) -- (1.83,1.85);
        \draw[red, >=stealth, ->] (-1.93,-1.95) -- (-1.68,-1.88);
	\end{tikzpicture}
	\put(-80,6){ \color{white}{\bf \small{RMSE:28.3HU}}}
	\put(-115,105){ \color{white}{\bf \small{Boosting}}} 
 
	\begin{tikzpicture}
         \begin{scope}
		[spy using outlines={rectangle,green,magnification=2,size=9mm, connect spies}]
		\node {\includegraphics[width=0.24\textwidth]{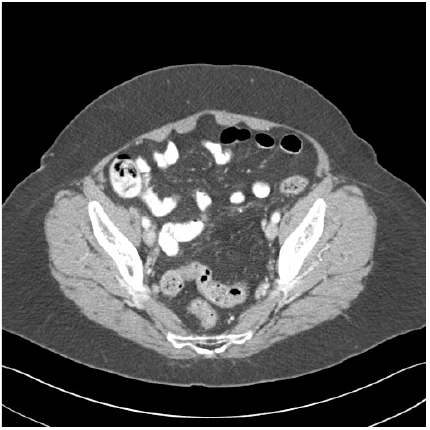} };
            \spy [green, draw, height = 0.77cm, width = 0.77cm, magnification = 2,
		connect spies] on (1.36,-0.5) in node [left] at (2,1.65);
  		\spy [green, draw, height = 0.77cm, width = 0.77cm, magnification = 2,
		connect spies] on (-0.57,-0.93) in node [left] at (-1.25,-1.65);
        \end{scope}
        \draw[red, >=stealth, ->] (2.02,2.02) -- (1.83,1.85);
        \draw[red, >=stealth, ->] (-1.93,-1.95) -- (-1.68,-1.88);
	\end{tikzpicture}
	\put(-80,6){ \color{white}{\bf \small{RMSE:0HU}}}
	\put(-115,105){ \color{white}{\bf \small{Reference}}} 
	\vspace{-0.1in}
	\caption{Reconstruction of slice 150 from patient L310 using various methods. The display window is [800, 1200] HU.}\label{Resultscomparision}
\end{figure*}

{\color{black}
Figures~\ref{fig:rmsebox}, \ref{fig:ssimbox}, and \ref{fig:snrbox} are box plots of RMSE, SSIM, and SNR values of (test) reconstructions with PWLS-EP, PWLS-ULTRA, PnP-ADMM, the standalone FBPConvNet, SS-FCN-ULTRA, PS-FCN-ULTRA, PS-FCN-PnP and Boosting. Figure~\ref{fig::RMSEevolution} shows the convergence curve of RMSE metric of PS-FCN-ULTRA, PS-FCN-PnP and Boosting methods.
We observe that the proposed parallel SUPER methods, namely PS-FCN-ULTRA and PS-FCN-PnP present lower RMSE values, higher SSIM values and higher SNR values in test slices than standalone MBIR and deep learning methods such as PWLS-EP, PWLS-ULTRA, PnP-ADMM, and FBPConvNet. Specifically, the SUPER methods can effectively handle highly damaged scans for which the performance of standalone supervised or unsupervised methods may be limited.
Compared with the deep boosting methods, the parallel SUPER learning methods show better mean RMSE, SSIM and SNR and narrower interquartile ranges.
Compared with the serial SUPER method, the parallel structure also stands as a better method with narrower interquartile ranges, reduced gap between the maximum and minimum RMSE values, SSIM values and SNR values. 

In our study, we have tried two unsupervised methods: the PWLS-ULTRA method and the PnP-ADMM method.
We can find that the PS-FCN-ULTRA and PS-FCN-PnP both achieve better RMSE than the standalone PWLS-ULTRA and PnP-ADMM methods.
We also observe that the PS-FCN-ULTRA method outperforms the PS-FCN-PnP method.
This improvement can be attributed to greater sophistication of ULTRA solver compared to that of the PnP-ADMM method. Furthermore, the standalone PWLS-ULTRA method outperforms the standalone PnP-ADMM method. Consequently, enhancing the performance of individual modules in the parallel SUPER model leads to an overall improvement in the parallel SUPER model's performance.

The superiority of the parallel SUPER method over baselines can also be illustrated by reconstruction quality metrics shown 
in Table~\ref{tab:results}, where we show the mean RMSE, SSIM and SNR values over all aforementioned methods.

The PS-FCN-ULTRA method attains the best RMSE, SSIM, and SNR values. Both PS-FCN-ULTRA and PS-FCN-PnP outperform the standalone FBPconvnet, PWLS-ULTRA, and PnP-ADMM schemes. In particular, the PS-FCN-ULTRA shows 7.2 HU and 10.4 HU lower RMSE and 2.4 dB and 2.8 dB higher SNR than the standalone supervised (FBPConvNet) and unsupervised (PWLS-ULTRA) methods, respectively. PS-FCN-PnP shows 6.2 HU and 12.4 HU lower RMSE and 2.1 dB and 3.2 dB higher SNR than the standalone supervised (FBPConvNet) and unsupervised (PnP-ADMM) methods, respectively. When compared with the SS-FCN-ULTRA method, the PS-FCN-ULTRA can achieve 3 HU lower RMSE and 0.9 dB higher SNR with the same supervised module and unsupervised module setup.

\subsection{Visual Results and Comparisons}

Figure~\ref{Resultscomparision} shows the reconstructions of a test example (slice 150 from patient L310) using all aforementioned methods.
The PWLS-ULTRA and PnP-ADMM methods effectively suppress noise and recover details, but may introduce blurry artifacts. The standalone FBPConvNet method removes noise and streak artifacts but introduces artificial features. While the deep boosting method improves upon FBPConvNet, it still has streak artifacts and blurred details, such as the upper right zoom-in area. The SS-FCN-ULTRA method enhances image quality, but some features are missing. The proposed Parallel SUPER methods (PS-FCN-ULTRA and PS-FCN-PnP) further improve reconstruction by removing noise and artifacts while preserving fine details, such as the line features in the lower left zoom-in area.
\begin{figure*}[!t]
	\centering  
  		\begin{tikzpicture}
		[spy using outlines={rectangle,green,magnification=2,size=9mm, connect spies}]
		\node {\includegraphics[width=0.24\textwidth]{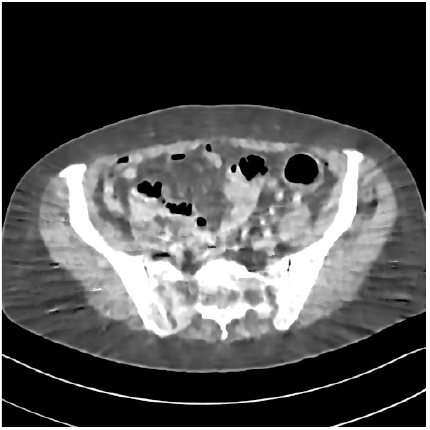} };
\spy [green, draw, height = 0.77cm, width = 0.77cm, magnification = 2,
		connect spies] on (0.5,-0) in node [left] at (1.95,1.6);
	\end{tikzpicture}
	\put(-79,6){ \color{white}{\bf \small{RMSE:33.8HU}}}
	\put(-117,108){ \color{white}{\bf \small{Layer 1}}} 	
 		\begin{tikzpicture}
		[spy using outlines={rectangle,green,magnification=2,size=9mm, connect spies}]
		\node {\includegraphics[width=0.24\textwidth]{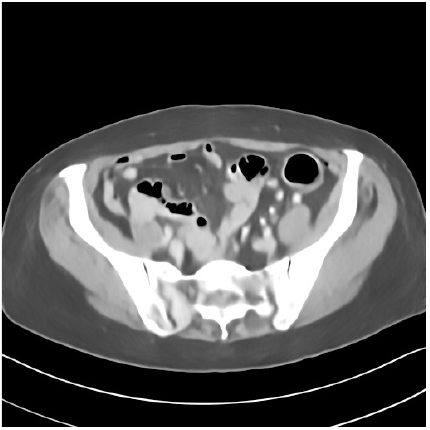} };
\spy [green, draw, height = 0.77cm, width = 0.77cm, magnification = 2,
		connect spies] on (0.5,-0) in node [left] at (1.95,1.6);
	\end{tikzpicture}
	\put(-79,6){ \color{white}{\bf \small{RMSE:27.2HU}}}
	\put(-117,108){ \color{white}{\bf \small{Layer 8}}} 	
	\begin{tikzpicture}
		[spy using outlines={rectangle,green,magnification=2,size=9mm, connect spies}]
		\node {\includegraphics[width=0.24\textwidth]{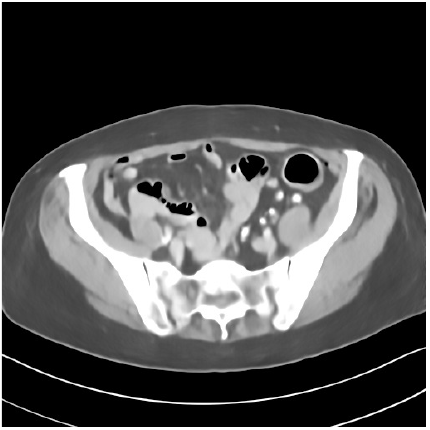} };
  \spy [green, draw, height = 0.77cm, width = 0.77cm, magnification = 2,
		connect spies] on (0.5,-0) in node [left] at (1.95,1.6);
	\end{tikzpicture}
	\put(-79,6){ \color{white}{\bf \small{RMSE:23.2HU}}}
	\put(-117,108){ \color{white}{\bf \small{Layer 15}}} 	
  \begin{tikzpicture}
		[spy using outlines={rectangle,green,magnification=2,size=9mm, connect spies}]
		\node {\includegraphics[width=0.24\textwidth]{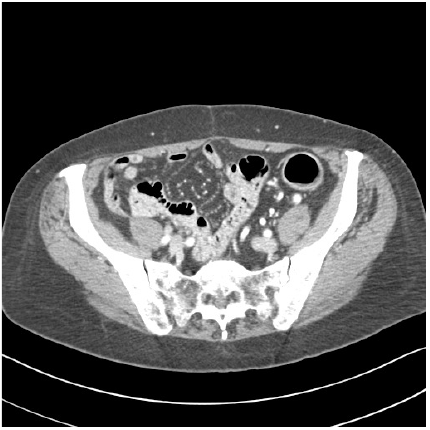} };
   \spy [green, draw, height = 0.77cm, width = 0.77cm, magnification = 2,
		connect spies] on (0.5,-0) in node [left] at (1.95,1.6);
	\end{tikzpicture}
	\put(-70,6){ \color{white}{\bf \small{RMSE:0HU}}}
	\put(-117,108){ \color{white}{\bf \small{Reference}}} 	
	\vspace{-0.1in}
	\caption{Image (slice 150 of patient L192) evolution over Parallel SUPER layers using the PS-FCN-ULTRA method. RMSE values are also indicated. The display window is [800, 1200] HU.}\label{Imageevolution}
\end{figure*}
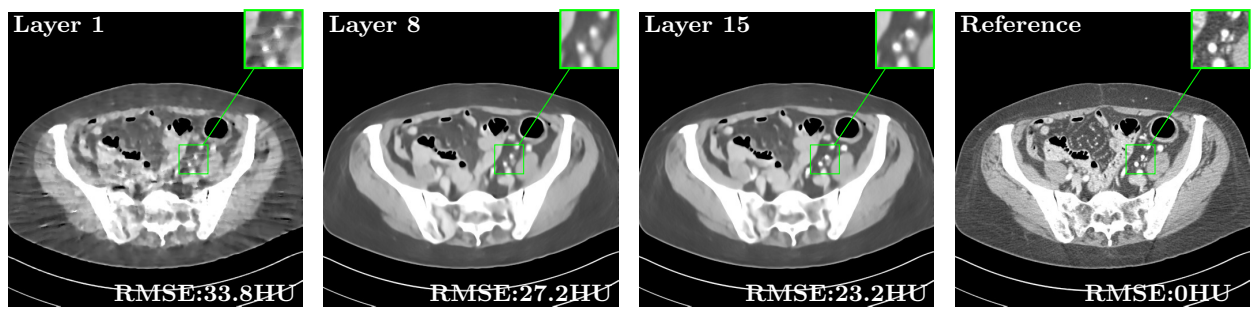

Figure~\ref{Imageevolution} illustrates the image evolution over layers in the parallel SUPER pipeline (i.e., passing through the stacked parallel SUPER blocks in the iterative reconstruction process) for  slice 150 of patient L192, when using the PS-FCN-ULTRA method. 
In early parallel SUPER blocks, the proposed SUPER-FCN-ULTRA method mainly removes noise and artifacts, while later SUPER layers mainly reconstruct details such as the zoom-in areas.
We can find that the image quality improves as the parallel SUPER blocks go deeper, where there are less blurred regions and clearer bone details.

\subsection{Effect of Regularization Parameter $\mu$ in Parallel SUPER Models}
We study the impact of the regularizer weight $\mu$ in the final block of parallel SUPER pipeline on reconstruction performance. Recall the optimization formulation in the unsupervised module \eqref{eqn::unsupervised_obj}. 
Setting $\mu=0$ let the regularization term that penalizes the discrepancy between output from the unsupervised module and output from the previous adjacent parallel SUPER block vanish. In this case, the unsupervised solver is independent from prior information from the SUPER pipeline. $\mu>0$ propels the unsupervised module to be aligned with the most recent parallel SUPER block output and this convex penalty term improves the stability condition of the optimization problem. 

We have observed that the unsupervised module can be unstable in its output quality as it can be sensitive to the initialization of the optimization process and hyperparameter setup. After multiple Parallel SUPER blocks process the input LDCT image, with $\mu=0$, the unsupervised module in later blocks could have degraded performance in denoising and consequently drive the combination weight $(1 - \lambda)$ assigned to unsupervised module to low values. Setting $\mu>0$ helps maintain the meaningfulness of the contribution from unsupervised modules in deep Parallel SUPER blocks.
Therefore, we choose $\mu = 5 \times 10^{5}$ to make the unsupervised module function as a non-trivial component.

Figure~\ref{lambdaevolution} shows the optimal sequence of parameter $\lambda$ with respect to different choices of parameter $\mu$ across SUPER blocks of the PS-FCN-ULTRA method.
When $\mu = 5\times 10^{5}$, the optimized parameter $\lambda$ in the first block is $\lambda = 0.9$, as limited iteration budget for unsupervised solver impedes its standalone performance and correspondingly the supervised module gains more weight. Optimized value of $\lambda$ drops to 0.25 in the following block as the PWLS-ULTRA solver performs stronger when applied to images with low levels of noise. The value of $\lambda$ increases in the subsequent blocks of the parallel SUPER method and converges to around 0.75 after 10 blocks. The convergence of $\lambda$ indicates both methods contribute to the reconstruction process and approach a balance between denoising ability and image detail preservation.

Larger values of $\mu$ leads to the greater impact of unsupervised PWLS-ULTRA solver in deep blocks of the parallel SUPER pipeline, as the final value to which the $\lambda$ sequence converges becomes smaller, despite supervised module gaining greater weight in the second parallel SUPER block. We point out that the unsupervised solvers in later blocks do not alter their input much with large $\mu$, which explains the reduced weight assigned to unsupervised modules.

\begin{figure}[H]
	\centering
		\includegraphics[height=5cm]{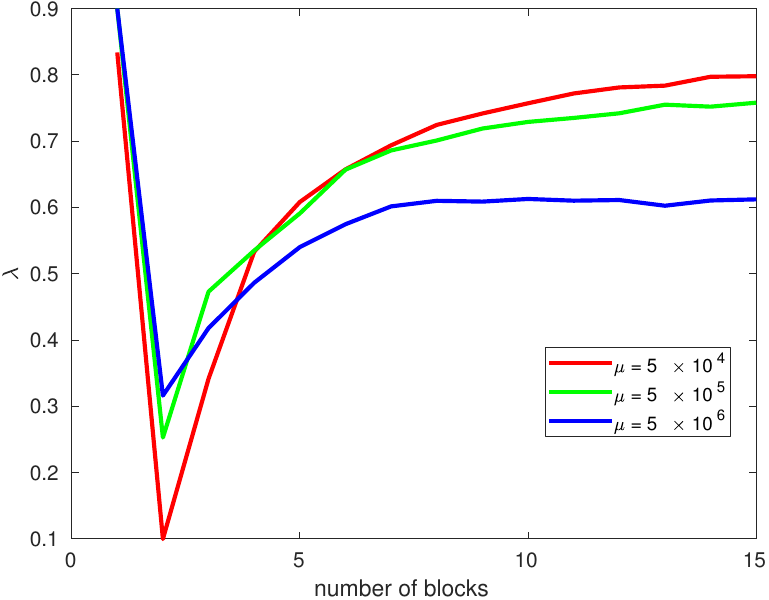}
	\caption{Parameter $\lambda$ 
 (over 500 training slices) across blocks in the PS-FCN-ULTRA method with different parameter $\mu$.}\label{lambdaevolution}
\end{figure}

\subsection{Effect of Regularization Parameter $\beta$ in Parallel SUPER Models}
We investigate the impact of the unsupervised regularization parameter $\beta$ on parallel SUPER by taking the PS-FCN-ULTRA method as an example. 
We fix the regularization parameter $\mu=5\times 10^5$ and use different values of the parameter $\beta$ for the PS-FCN-ULTRA method. 
Figure~\ref{beta_parameter} shows mean RMSE results of PS-FCN-ULTRA using different $\beta$ values during training and testing over 20 test slices.
When no unsupervised learning-based prior is involved during training ($\beta=0$), the reconstruction quality deteriorates significantly. 
When we set $\beta=5\times 10^3$, we achieve around 18~HU RMSE improvement compared to using $\beta=0$. As figure~\ref{beta_parameter} indicates, larger $\beta$ leads to worse RMSE values, thus making $\beta=5\times 10^3$ a locally optimal choice in the neighbourhood.
\begin{figure}[H]
	\centering
		\includegraphics[height=5cm]{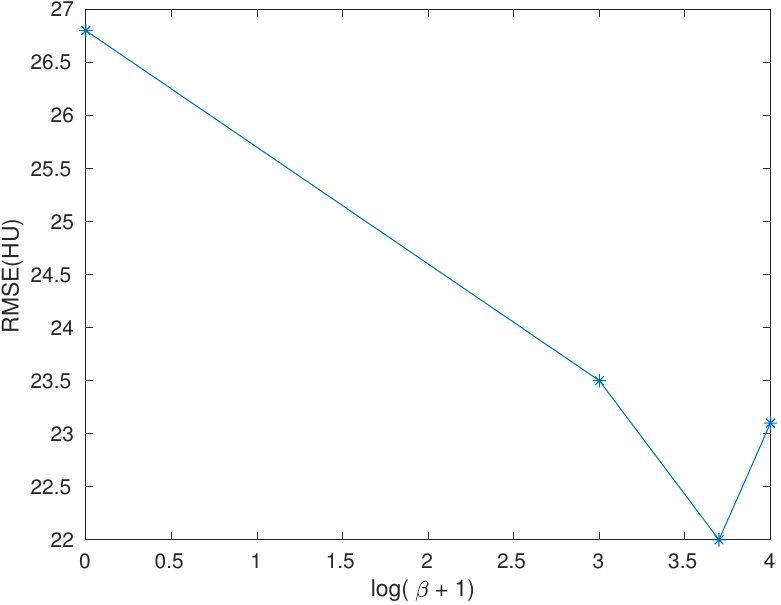}
	\caption{Mean RMSE (HU) of $20$ test slices using different $\beta$ values in the PS-FCN-ULTRA method.}\label{beta_parameter}
\end{figure}

\section{Conclusion}
This study proposes a Parallel SUPER learning framework for combining supervised and unsupervised methods for low-dose X-ray CT image reconstruction.
The proposed Parallel SUPER learning framework combines physics-based forward models, statistical models of measurements and noise, machine learned models, and analytical image models in a common framework. 
We study two example Parallel SUPER learning methods by combining the FBPConvNet architecture for the supervised learning-based mapping, and the union of learned transforms model and the PnP-ADMM method for the unsupervised scheme.
We have found that the proposed Parallel SUPER model helps improve the reconstruction quality in terms of reducing noise and artifacts and reconstructing structural details better compared with the standalone supervised module, unsupervised module, or the prior Serial SUPER model.
For future work, we plan to investigate a general formulation that combines the supervised module and unsupervised module using diverse structures. We also aim to explore the convergence theory of the proposed method.

\bibliographystyle{alpha}
\bibliography{mybib}

\end{document}